\documentclass[english,pdftex,superscriptaddress,prb,twocolumn,aps,showkeys,longbibliography]{revtex4-1}
\usepackage[T1]{fontenc}
\usepackage[latin9]{inputenc}
\usepackage{float}
\usepackage{textcomp}
\usepackage{amsmath}
\usepackage{amssymb}
\usepackage{graphicx}
\usepackage{subscript}

\makeatletter
\@ifundefined{textcolor}{}
{%
 \definecolor{BLACK}{gray}{0}
 \definecolor{WHITE}{gray}{1}
 \definecolor{RED}{rgb}{1,0,0}
 \definecolor{GREEN}{rgb}{0,1,0}
 \definecolor{BLUE}{rgb}{0,0,1}
 \definecolor{CYAN}{cmyk}{1,0,0,0}
 \definecolor{MAGENTA}{cmyk}{0,1,0,0}
 \definecolor{YELLOW}{cmyk}{0,0,1,0}
}

\usepackage{color}\usepackage{babel}

\usepackage{babel}

\usepackage{babel}

\makeatother

\usepackage{babel}
\begin{document}

\title{High-quality photonic crystals with a nearly complete band gap obtained
by direct inversion of woodpile templates with titanium dioxide }

\author{Catherine Marichy} \thanks{These authors contributed equally to this work}

\author{Nicolas Muller} \thanks{These authors contributed equally to this work}

\author{Luis S. Froufe-P\'erez}

\author{Frank Scheffold}

\email{frank.scheffold@unifr.ch}

\affiliation{Department of Physics, University of Fribourg, Chemin du Musée 3,
CH-1700, Fribourg, Switzerland}

\date{\today}

\renewcommand{\abstractname}{\vspace{-\baselineskip}}

\begin{abstract}
\bigskip
\bigskip
{Photonic crystal materials are based on a periodic modulation
of the dielectric constant on length scales comparable to the wavelength
of light. These materials can exhibit photonic band gaps; frequency
regions for which the propagation of electromagnetic radiation is
forbidden due to the depletion of the density of states. In order to exhibit
a full band gap, 3D PCs must present a  threshold refractive
index contrast that depends on the crystal structure. In the case
of the so-called woodpile photonic crystals this threshold is comparably low, approximately
1.9 for the direct structure. Therefore direct or inverted woodpiles made of high refractive index
materials like silicon, germanium or titanium dioxide are sought after.
Here we show that, by combining multiphoton lithography and atomic layer deposition, we can achieve a direct inversion of polymer templates into TiO$_{2}$ based
photonic crystals. The obtained structures show remarkable optical
properties in the near-infrared region with almost perfect specular
reflectance, a transmission dip close to the detection limit and a
Bragg length comparable to the lattice constant. } 
\end{abstract}


\maketitle

Even though three-dimensional woodpile Photonic crystals (PCs), made of high index materials such as titanium
dioxide\citep{Subramania_Adv_Mat_2010_A} and silicon\citep{Subramania_Adv_Mat_2010_B},
have been fabricated for instance by successive sputtering and electron
beam patterning, the infiltration of a lithographically structured polymer template appears to
be a more useful strategy for obtaining large three dimensional PCs \cite{Hermatschweiler_Adv_Func_Mat_2007,Staude_Opt_Lett_2010}
and for embedding functional subunits for application in optical circuits
or devices \cite{joannopoulos2011photonic,rinne2008embedded,Staude_Opt_Lett_2011}. Tétreault et al. reported
the first silicon replica of such a woodpile template using the silicon double
inversion technique \citep{Tetreault_Adv_Mat_2006}. A SiO\textsubscript{2} inverse woodpile PC
was used as an intermediate structure that was subsequently infiltrated
with silicon by employing high temperature chemical vapor deposition (CVD).
Later, the same group reported on the fabrication of silicon inverse
woodpile PCs\citep{Hermatschweiler_Adv_Func_Mat_2007}. In this case,
a SiO\textsubscript{2} shell was deposited around the polymer rods
of the template to preserve the log-pile structure during the silicon
deposition process. By these and related approaches, silicon hollow-rod
woodpile PCs\citep{Gratson_Adv_Mat_2006}, waveguides\citep{Staude_Opt_Lett_2011}, hyperunifrom \cite{muller2014silicon}
and complete band gap PCs at telecommunication wavelengths
around 1.55 $\mu$m\citep{Staude_Opt_Lett_2010} have been obtained.
Another well-known high refractive index material that has been widely  used in the past is titania (TiO\textsubscript{2}, titanium dioxide)\cite{wijnhoven1998preparation,Subramania_Adv_Mat_2010_A,Frolich_Adv_Mat_2013}.
The advantages of titania, compared to silicon, are its transparency
in the visible to mid-infrared region as well as the possibility to
chemically wet-process. In addition, titania can be deposited at moderate
temperatures around or below 100 $^{\circ}$C. This makes titania
a particularly well-suited material to infiltrate threedimensional
polymeric scaffolds directly without the need for any additional infiltration
steps which tend to lead to structural deterioration.
\newline In this work we present the fabrication, structural and optical characterization
of titania hollow-channel and inverse woodpile PCs.
Our work reports two main achievements in the field. First, we report on the fabrication of high-index woodpile
PCs in a single infiltration step process. This facilitates,
as we will show, the fabrication of structures and yields an outstanding
structural integrity and surface quality. Second, we present a direct
measurement of the Bragg length L\textsubscript{B} in our high-refractive-index
PCs. The Bragg length is a key parameter for the characterization
of photonic band gap materials as it sets the length scale an evanescent
wave can penetrate into a PC material. As such it also sets the lower
bound for the design of integrated structural features such as bends
and cavities. The latter have to be separated by several Bragg lengths
in order to prevent cross-talkink or tunneling. Despite the importance
of the Bragg length very little experimental data on high-refractive-index materials has been reported to date in the literature. A number of studies have studied the Bragg length for the case of low-index materials in the vicinity of the corresponding pseudo-gaps \cite{vlasov1997existence,galisteo2003optical,Garcia_PRB_2009}. Here
we demonstrate that owing to the precision of our materials fabrication
process we can reproducibly fabricate high-refractive-index photonic crystals with different
layer thicknesses ranging from 12 to 32 layers (one unit cell is composed
of four layers). By measuring the transmission dip minimum as a function
of the thickness of the PC layer, we directly determine
the Bragg length L\textsubscript{B} for one given lattice orientation.
\newline Photonic titanium dioxide structures have already been realized in
the past by single or double inversion/infiltration methods of polymer templates
by sol-gel wet processing\citep{Heinroth_J_Mat_Sci_2009,Lee_APL_2010,Park_Langmuir_2013,Serbin_OPEX_2004},
sputtering\citep{Subramania_OPEX_07} and via atomic layer deposition
(ALD)\citep{Biswas_JOSA_B_2005,Frolich_Adv_Mat_2013,Lee_APL_2007,George_Chem_Rev_2010,Knez_Adv_Mat_2007,Leskela_Mat_Csi_Eng_C_2007,Marichy_Adv_Mat_2012,Pinna_book_2011,Gaillot_book_2011}.
ALD provides control of the film thickness at the atomic scale
already at moderate temperatures\citep{George_Chem_Rev_2010,Knez_Adv_Mat_2007}.
Consequently, ALD also allows the elaboration of a thin protective
shell around thermally unstable polymer templates for example by depositing
Al\textsubscript{2}O\textsubscript{3}\citep{Santamaria_Adv_Mat_2007,Shir_JVSTB_2010}
or TiO\textsubscript{2}\citep{Muller_Adv_Mat_2014}. In a subsequent
step such stable hybrids can be infiltrated with silicon using
thermal CVD as shown for the case of polymer woodpiles\citep{Santamaria_Adv_Mat_2007,Shir_JVSTB_2010}
and hyperuniform network structures\citep{Muller_Adv_Mat_2014}. While
woodpile PCs made of a polymer core and of inorganic shell have also
been obtained using TiO\textsubscript{2 }ALD coating\citep{Biswas_JOSA_B_2005,Lee_APL_2007},
a direct full inversion of polymer woodpile PCs with ALD of TiO\textsubscript{2}
has not been reported so far to our knowledge. Graugnard et al.\citep{Graugnard_APL_2009}
reported TiO\textsubscript{2} replica of holographically defined
polymer templates by double inversion techniques using ALD. After inversion of
the polymer structure into Al\textsubscript{2}O\textsubscript{3}
using a low temperature ALD process, a second ALD infiltration with
TiO\textsubscript{2} was performed. The 3D PC consisted of 23 layers
and the successful infiltration by ALD through several layers was
demonstrated. Recently, direct TiO\textsubscript{2} woodpile PCs
were obtained by a similar approach using an intermediate ALD ZnO
step\citep{Frolich_Adv_Mat_2013}. Despite the comprehensive work
that has already been performed on TiO\textsubscript{2} woodpile-based
3D PCs, very few of these studies show evidence for the theoretically
predicted near-zero transmission dips as well as near 100\% reflection
peaks. This suggests that the material properties, such as mean refractive
index and structural integrity, are not sufficiently close to the
assumptions made in theoretical band structure calculations.

\begin{figure*}
\centering\includegraphics[width=0.8\textwidth]{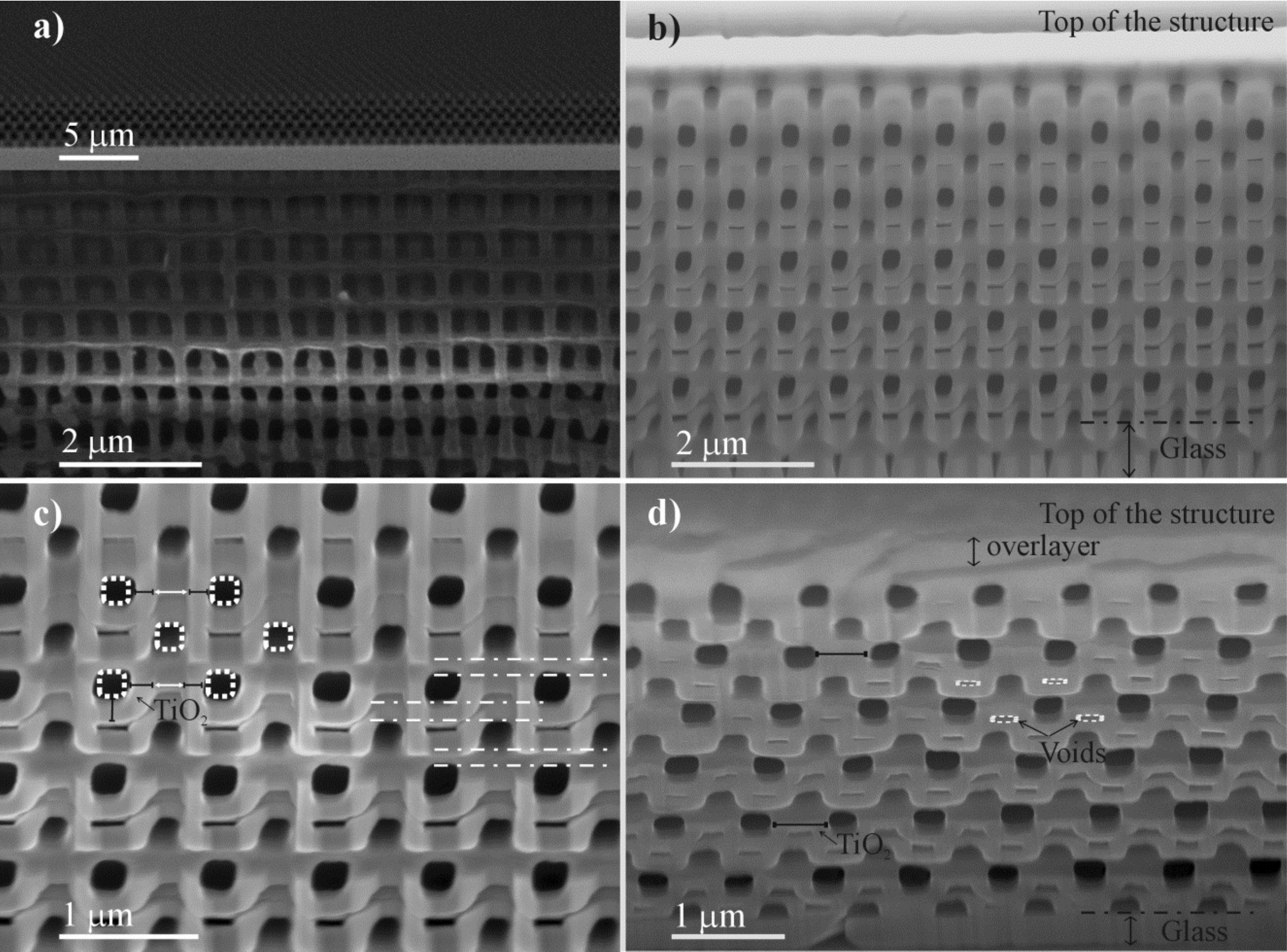} \caption{\label{fig_1}SEM images of a) polymer, b,c) partially TiO$_{2}$
infiltrated woodpile structures, and d) nearly completely infiltrated
structures obtained after 1500 and 4500 ALD cycles, respectively.
The polymer template has been removed by calcination b,c,d). Cross-sections
are cut using a focus ion beam with viewing angle varying from 25\textdegree{}
d) to 45\textdegree{} b), c). In the enlarged image c) of the partially
TiO$_{2}$ infiltrated woodpile shown in b), the dotted circles and
dashed lines represent the rod positions in the initial polymer template;
the TiO$_{2}$ is indicated by black lines while the interspaces between
two hollow TiO$_{2}$ hollow channels are highlighted by the full
white lines. The image in d) reveals the presence of air voids (dotted
circles) inside the completely inverted titanium dioxide structure.
The metal oxide over-layer on the top of the structure is also shown.}
\end{figure*}

\section{RESULTS}
\subsection*{Material fabrication and characterization} 
\begin{figure*}
\centering\includegraphics[width=0.8\textwidth]{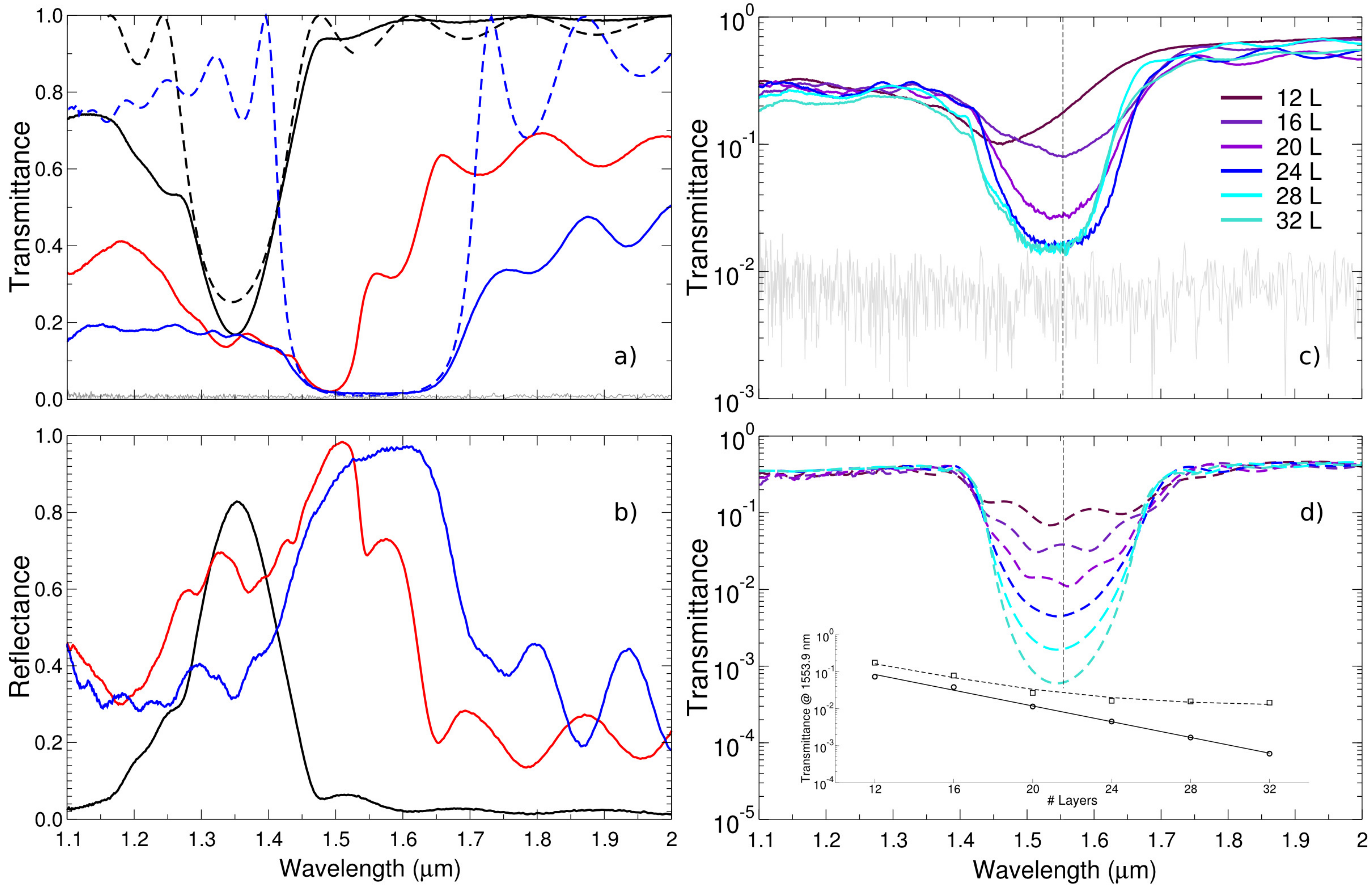} \caption{\label{fig_2} a) Transmittance and b) reflectance spectra from
the polymer template as obtained by DLW (black full line), the partially
infiltrated TiO\textsubscript{2} hollow-rod structures (red full
line) and the TiO\textsubscript{2} inverted structure (blue full
line). Calculated transmittance spectra at normal incidence are shown
as dashed lines. c) Measured transmittance spectra in log-plot representation
for TiO\textsubscript{2} inverse woodpiles with different numbers
of layers (see legend). The gray line indicate the background noise
signal recorded on a gold mirror as a beam block. d) Normal incidence
calculated spectra with a gap center wavelength of 1.55 $\mu$m (dashed
vertical line). Inset in d) shows the calculated (circles) and measured (squares)
transmittance at a wavelength lying in the ${\Gamma-X'}$ bandgap.
The full line shows the best fit of the simulated data to an exponential
$T=1.6exp(-0.245\times\#Layers)$. The dashed line shows an exponential
decay with a slightly adjusted prefactor using same decay length as
in the calculations leading to excellent agreement with the experimental
data $T=2.9exp(-0.245\times\#Layers)+0.012$. A small baseline value
is added for a best fit in order to take the instrument background
noise and other imperfections of the measurementinto account.}
\end{figure*}

In the present work we show that direct infiltration with ALD and
subsequent removal of the polymer template by calcination result in
high-quality photonic materials that allow a near-quantitative comparison
between calculated and experimental spectra of TiO\textsubscript{2}
hollow-rod and inverse woodpile PCs. Combining multiphoton lithography, also known as direct laser writing
(DLW), with atomic layer deposition, we successfully invert polymer
woodpiles templates into TiO\textsubscript{2}. Due to the high conformity and the uniformity of the infiltration
process, high-quality TiO\textsubscript{2} inverse woodpiles are
obtained. The latter display remarkable optical properties in the near-infrared
region (1.4-1.7 $\mu$m) with a reflectance peak close to 100 \% and a
transmittance dip close to the detection limit.
\newline The polymer woodpile templates, shown in Figure \ref{fig_1}a, are
written using DLW and the novel dip-in method combined with a shaded-ring
filter using the commercial \quotesinglbase Photonic Professional`
platform (Nanoscribe GmbH, Germany). This particular combination of
techniques permits to fabricate rods with a low aspect ratio and thus
leads to structures with a unprecedented quality. Indeed, while
by regularDWL stop bands with less than 60 \% in reflection are typically
measured with unpolarized light\citep{Hermatschweiler_Adv_Func_Mat_2007,Deubel2004},
the initial PCs presented here exhibit a band gap of 80-85 \% in reflection
and a dip in transmission lower than 20 \% (Figure \ref{fig_2}).
The templates present on average a lateral rod spacing of $\sim$ 750 nm and a rod diameter of 235 nm yielding  stop bands around
1.2-1.35 $\mu$m. Only small variations are observed between different
fabrication runs. Massive polymer walls of 10 $\mu$m thickness are
lithographically placed around the woodpiles in order to increase
their mechanical stability along the development and infiltration
processes. Structures made of 24 layers are partially and completely
infiltrated with TiO\textsubscript{2 }by ALD at moderate deposition
temperatures. The degree of infiltration is controlled by the number
of applied cycles. We subsequently remove the polymer by high temperature
calcination and transform the as-deposited amorphous titanium dioxide
into its denser anatase phase. Incomplete infiltration with titania
thus leads to structures composed of hollow titania rods and for longer
infiltration times we obtain completely inverted titania woodpile
PCs. In Figure \ref{fig_1}b-d, scanning electron microscope (SEM)
images recorded from cross-sections of both types of TiO\textsubscript{2}
structures reveal their high-quality and the good homogeneity of the
deposition from top to bottom without significant variations of the
thickness. The bright contrast indicates the TiO\textsubscript{2}
and the dark contrast the voids left over by the degraded polymer.
One should note that the writing process is started in a virtual depth
of 2 layers inside the glass substrate (shown in Figure \ref{fig_1}b,d)
to guarantee a continuous laser writing process along the axial direction
which is necessary to ensure the adhesion of the polymer template
to the substrate. Woodpile PCs composed of hollow rods are obtained
by partial infiltration (stopped after 1500 cycles). In Figure \ref{fig_1}c, a titania layer of finite
thickness (indicated by black lines) is grown \ around the polymer
rods (their initial positions are pointed out by dotted white circles).
Moreover the quincunx distribution of out- and in-plane layers is
clearly visible. The in-plane layers are highlighted by dotted white
lines. Horizontal interspaces between two neighboring shells of an
out of plane layer can be distinguished and indicate the absence of
coalescence of the shells due to partial infiltration. In the case
of the fully infiltrated TiO\textsubscript{2} inverse woodpile no
such interspaces are noted and an over-layer is visible on top of the
structure (Figure \ref{fig_1}d). The infiltration, occurring in a
conformal manner, unavoidable small voids are formed due to clogging
that prevents the precursors to diffuse further in the final stages
of the infiltration process.

Atomic force microscopy (AFM) measurements (Supplementary
Fig. 1) reveal the smoothness of the surface with variations of the root mean square
roughness (RMS) along a rod on the order of 2-3  nm only which is a sign of the high conformity of the coating. In
the AFM images of the hollow rod woodpile PCs the first layer perpendicular
to the top one is visible while only the top rods are noted for the
TiO\textsubscript{2} inverse structure, confirming the partial and
full infiltration.  Spectroscopy data in the mid-IR confirm the removal of the polymer during the high temperature annealing step, and the densification of the TiO\textsubscript{2} (Supplementary
Fig. 2).
\subsection*{Near infrared spectroscopy}
The optical properties of the structures are characterized by means of a
Fourier transform infrared spectrometer in combination with a grazing
angle objective. We measure the near-infrared reflectance and transmittance along incident angles which are contained in a cone ranging from
10$^{\circ}$ up to 30$^{\circ}$ with respect to the surface normal. The
normalization of the transmittance and reflectance are performed on
a bare glass substrate and a gold mirror, respectively. Figure \ref{fig_2}
shows the optical response of polymer, TiO\textsubscript{2} hollow-rod
and TiO\textsubscript{2} inverse woodpile PCs in transmittance and
in reflectance. For the polymer woodpile PCs (full black lines) we observe
a strong and narrow transmission dip as well as a strong reflection
peak at 1.35 $\mu$m. After both partial and complete infiltration
and polymer removal, a redshift as well as a broadening of the dip
and peak are noticed. For the case of completely infiltrated structures, the transmittance
drops below 2 \% and the reflectance reaches almost 100 \%. The inverse
woodpile PCs (blue full line) show a robust stop band centered at 1.55
$\mu$m which is 18.5 \% wide, while the band gap at 1.5 $\mu$m of
the hollow-rod woodpile is less well defined even though still relatively
broad (see also Supplementary
Fig. 3). As visible in Figure \ref{fig_2}a, the transmission characteristics
of the TiO\textsubscript{2} hollow rod-structure (red full line)
are not as well developed as for the inverse woodpile. For both titanium dioxide PCs, Fabry-Pérot fringes are visible, highlighting the
good quality of the infiltration and the integrity of the structures
during the entire fabrication process. Indeed, only high-quality PCs
with homogeneous optical properties along the z-axis can exhibit such
features.
\subsection*{Comparison to numerical calculations}
\begin{figure*}
\centering\includegraphics[width=0.8\textwidth]{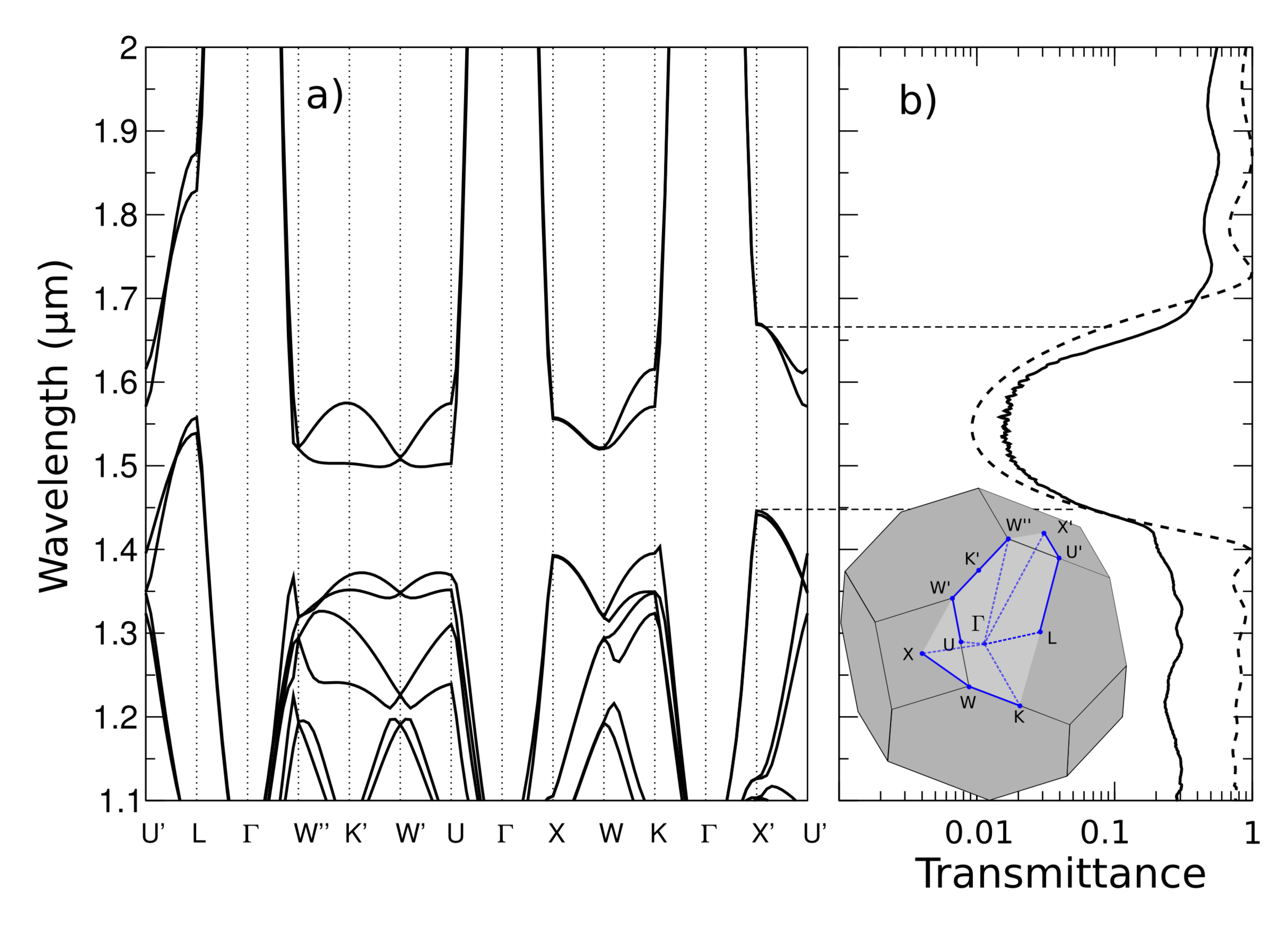} \caption{\label{fig_3} a) Calculated photonic band diagram for the inverted
woodpile PC shown in Figure \ref{fig_1}d. The parameters of the band
structure calculation are taken from the comparison to the experimental
data as shown in panel b). In order to take the effect of voids in
the structure into account, a reduced effective refractive index was
determined to n\textsubscript{eff}=2.12 instead of n=2.45 for bulk
TiO\textsubscript{2}-anatase. b) Calculated normal incidence transmission
spectrum (angle integrated at the output) for an inverted woodpile
of 24 layers (dashed line) and the comparison to the measured spectrum
(full line). The inset shows the geometry of the first Brillouin zone,
a slightly stretched FCC lattice, as well as the points of high symmetry. }
\end{figure*}

Comparison with numerical calculations shows good agreement with the
experiments. The dashed lines in Figures \ref{fig_1}a indicate the total
transmission spectra obtained from Finite Differences Time Domain
(FDTD) simulations\citep{Oskooi_Comp_Phys_Comm_2010}. To obtain suitable
parameters for modeling the woodpile structure, photonic band structure
calculations were first carried out in order to fit the $\Gamma-X'$stop
gap center and width (corresponding to normal incidence) to the measured
spectra. As shown in Figure \ref{fig_2}, the position of the measured
band gap, its depth, and the Fabry-Pérot oscillations are nearly quantitatively
reproduced by the calculations.
\newline In order to investigate the influence of the structure height on the
stop band, we fabricate woodpile PCs templates of different heights
with a number of layers ranging from 12 to 32. Subsequently, the templates
are fully infiltrated with ALD TiO\textsubscript{2} and calcinated.
The number of ALD cycles has been reduced to 3500 in order to significantly
decrease the cap -layer. The transmittance spectra of the obtained
inverse woodpile PCs are presented in Figure \ref{fig_2}c. A minimum
of 24 layers is required to reach a transmittance as low as 1-2\%.
A further increase of the height of the PC does not reduce the apparent
depth of the gap. This observation can be explained by residual fabrication
errors and distortions during development, infiltration and polymer degradation.
However, the contribution of these imperfections cannot easily be
quantified since the FTIR spectroscopy as well as the normalization
can also lead to errors that may even become dominant for transmission
coefficients on the order of one percent as shown in figure \ref{fig_2}c
and d.
\newline Next we compare the experimental results to FDTD calculations at normal
incidence (Figure \ref{fig_2}d). Fitting the decay of T\textsubscript{Gap}
(L) from FDTD calculations at a wavelength $\lambda$=1553.9 nm with
an exponential (inset Figure \ref{fig_2}d), we again find a Bragg
length of L\textsubscript{B}$\sim$4 layerswhich corresponds to the
thickness of one unit cell. We define the Bragg length L\textsubscript{B}
as the transmittance decay length. Using this theoretically obtained
Bragg length and a baseline transmittance of 1.2 \% an excellent fit
to the experimental data is obtained. The difference in the prefactor
of the exponential can be attributed to the subtle differences in
the surface termination between the actual structure and the simulated
one. The experimental, quantitative extraction of the Bragg length
of a photonic material with a nearly complete bandgap is a remarkable result
of this study. Previous measurements of the Bragg length have been
reported for low-index polystyrene opal PCs  \cite{vlasov1997existence,galisteo2003optical,Garcia_PRB_2009}
and for silicon inverse woodpile PCs\citep{Huisman_PRB_2011}. In
the latter case a Bragg attenuation length of $\sim$7 layers was
found. It is telling that this value reported previously is actually
higher than the one measured in the present study despite the nominally
lower refractive index of bulk TiO\textsubscript{2}. We thus believe
that the small Bragg length observed in our study is yet another proof
for the excellent quality of our titania PC. Finally, in Figure \ref{fig_3},
we show the calculated full band diagram obtained for the effective
material properties extracted from the comparison of theory and experiment
in transmission and reflection. In order to take the effect of voids
in the structure into account a reduced effective refractive index
n\textsubscript{eff}=2.12 is used for the calculations instead of
n=2.45 for bulk TiO\textsubscript{2}-anatase. The band structure
displays a nearly complete band gap with a center wavelength of 1.55
$\mu$m. Only around the $U^\prime - L-\Gamma$ for narrow solid angle bands
slightly overlap. Figure \ref{fig_3}b shows again the excellent agreement
between experiment and numerical FDTD calculations for the case of
normal incidence transmission.

\section*{DISCUSSION}
We were able to invert polymeric woodpile PCs into titania by using
a single step infiltration approach. The high-quality and homogeneity
of the as-infiltrated structures is confirmed by SEM and the presence
of Fabry-Pérot fringes in the optical response. The PCs materials
fabricated exhibit very small values for transmission as well as a
near 100 \% reflection peak at a wavelength of 1.55 $\mu$m and a
gap width of 18.5 \%. The results are obtained reproducibly and thus provide a robust platform for
the future implementation of PC cavities and waveguiding. Thin air
channels inside the inverted structure can be attributed to the conformal
nature intrinsic to ALD processes and can be taken into account for
FDTD modeling by using a reduced effective refractive index of the
TiO\textsubscript{2}. The measured spectra are in good agreement
with band diagram calculations and transport simulations. We can clearly
distinguish between a partially and completely infiltrated case. Furthermore,
the experimental method presented here is highly flexible and can
be used to invert arbitrary two or three-dimensional polymeric structures
into titania.

\section*{METHODS}
\subsection*{Direct-Laser Writing (DLW)}
The woodpile photonic crystals were fabricated using a commercially
available setup (Photonic Professional from Nanoscribe GmbH) in combination
with the novel available Dip-In technique and a shaded ring filter.
Structures were written on glass and CaF\textsubscript{2 }(Crystan,
UK) substrates by \textit{dipping }the objective directly inside a
liquid negative-tone photoresist (IP-DIP, Nanoscribe, Germany) and
by fine-tuning the laser power. Two successive development baths in
PMMEA (propylene glycol monomethyl ether acetate) for 10 min and a
consecutive bath in isopropanol for 8 min were chosen. A gentle drying
of the structures is achieved by redirecting a stream of N\textsubscript{2}
through a bubbler containing isopropanol.

\subsection*{Titania Single-Inversion}
Woodpile PCs were subsequently infiltrated with TiO\textsubscript{2}
by atomic layer deposition (ALD). Depositions took place at 110 $^{\circ}$C
in a commercial ALD reactor (Savannah 100 by Cambridge Nanotech, Inc.)
operating in exposure mode. A Si wafer was coated simultaneously as
reference. Titanium isopropoxide (Sigma-Aldrich, 97 \% purity) and
DI water were used as metal and oxygen source, respectively, and introduced
alternately by pneumatic ALD valves. Their respective stainless steel
reservoirs were kept at 80 $^{\circ}$C and at room temperature. For
the deposition, pulse times of 0.05 s and 0.02 s were used for the
titanium precursor and oxygen source, respectively, under a carrier
gas flow of 5 sccm. The residence time without N\textsubscript{2}
flow and the purge time under 20 sccm of N\textsubscript{2} were
set to 20 s and 90 s, respectively. Considering a nominal growth per
cycle of 0.8 Å, the number of cycles was adjusted to yield partially
and completely infiltrated woodpile PCs. A further calcination at
600 $^{\circ}$C for 4 hours in a tubular furnace was performed to
remove the polymer and to convert the as-deposited amorphous TiO\textsubscript{2 }into
the denser anatase phase, which exhibits a refractive index of 2.45.
A heating ramp of 200 $^{\circ}$C per hour was chosen to reduce the
thermal deterioration of the structure during the calcination process.

\subsection*{Material Characterization}
The structure of a direct polymeric and TiO\textsubscript{2} inverted
woodpiles was studied with scanning electron microscopy (SEM), using
a Sirion FEG-XL30 S (FEI) microscope working between 5 and 10 kV.
Cross-sections were realized by focused ion beam using a Dualbeam
NOVA600 Nanolab (FEI). The etching was realized using a Ga ion beam
with an acceleration of 30 kV at a current of 3 nA for a couple of
minutes followed by a cleaning at 30 kV and 1 nA beam.
\newline The surface roughness and topography of the inverted TiO\textsubscript{2}
woodpile PCs were investigated by atomic force microscopy using a
NTEGRA Aura (NT-MDT) in topographic mode with a SKM tip (Supplementary
Fig. 1).

\subsection*{Optical Characterization}
Measurements of the optical response of the woodpile PCs were taken
using a Fourier transform IR interferometer (Bruker Vertex 70) coupled
to a microscope (Bruker Hyperion 2000, liquid N\textsubscript{2}-cooled
InSb detector). The objective was a 36x Cassegrain, numerical aperture
0.52. The transmittance of light incident at an angle between 10$^{\circ}$
and 30$^{\circ}$ to the surface normal was measured. Spectra were normalized
on the same substrate as the residing structures and on a gold mirror
for transmittance and reflectance measurements, respectively.

\subsection*{Numerical calculations of the PCs optical properties}
Calculations of the transmittance and reflectance spectra were performed
using the FDTD software package MEEP\citep{Oskooi_Comp_Phys_Comm_2010}.
Band structure calculations were carried out with the freely available
MIT photonic bands (MPB) software\citep{Johnson_OPEX_2001}.
\newline SEM micrographs, Figure \ref{fig_1}d, indicate that the air channels
in the inverted and thermally annealed structure show an almost rectangular
cross section. Thus for the calculations we assume a rectangular geometry
with a height of R\textsubscript{h}=415 nm and a width of R\textsubscript{w}=243
nm. For the in-plane channel separation we use a=651 nm and for the
inter-plane distance h=253 nm. As suggested previously\citep{Staude_Opt_Lett_2010},
the effect of the presence of voids in the inverted structure can
be taken into account by considering an effective refractive index.
\newline To numerically determine the transmission properties of the inverted
woodpile, transport simulations were performed with the freely available
MEEP software\citep{Oskooi_Comp_Phys_Comm_2010} on a finite stack
of a woodpile PC using the same geometrical parameters. The simulated
transmission spectra shown in the figures were obtained by calculating
the total transmittance of plane waves at normal incidence ($\Gamma-X'$)
(see also Supplementary
Fig. 3). The numerical comparison to
the experimental data was performed using a cell size of edge length
$\sim$15.6 nm and assuring that the energy is conserved at least
within 0.3 \% in all cases. Matching the position of the gap in the
$\Gamma-X'$ direction with the results obtained in transmission yields
an estimated effective refractive index n\textsubscript{eff}=2.12
instead of a nominal value of n=2.45 for bulk TiO\textsubscript{2}
in the anatase phase.
\newline The band structure is calculated based on these assumptions is shown
in Figure \ref{fig_3}a. In the inset of Figure \ref{fig_3}b, the
path in reciprocal space is shown in detail together with a sketch of the
first Brillouin zone of the lattice. As can be seen, the structure
corresponds to a slightly stretched Face Center Cubic (FCC) lattice.
Indeed, in the current case h/a=0.388, while in a perfect FCC lattice
h/a=0.353.
\newline To determine the transmission spectrum of the polymer woodpile templates
shown in Figure \ref{fig_2}, we assume elliptically shaped rods.
For the refractive index of the polymer we assume n=1.5. (The parameters
used in the numerical calculations are adapted for a best fit to the
experimental data.) From this we obtain a rod height R\textsubscript{h}=343
nm, rod width R\textsubscript{w}=323 nm, an in-plane rod separation
a=689 nm and an inter-plane distance h=267 nm. These lattice parameters
are slightly larger, by about 6 \%, compared to the inverted titania
structure. The exact origin of this small difference is not clear
as the exact choice of the (best)fitting parameters can be influenced by a
number of factors. First, the structure and the optical properties
of the material are coupled via the average refractive index of the
material. Second, during the  inversion and calcination process
small geometrical changes of the whole structure but also the rod/channel
cross-section can be observed. All of this can have an influence on the
optical parameters. Modeling such small imperfections is difficult
and beyond the scope of this work.

\section*{Acknowledgements}
The present project has been financially supported by the National
Research fund, Luxembourg (project No. 3093332), the Swiss National
Science Foundation (projects 132736 and 149867) and the Adolphe Merkle
Foundation.

\section*{Author contribution}
FS, CM and NM conceived the study. CM and NM contributed equally to the material
synthesis, fabrication and optical characterization
of the samples. LSFP performed the numerical simulations for band
structures and light transport. All authors contributed equally to
the writing of the manuscript.

\section*{Competing financial interests}  
The authors declare no competing financial interests.

\pagebreak
\widetext
\begin{center}
\textbf{\large Supplementary Figures: High-quality photonic crystals with a nearly complete band gap obtained
by direct inversion of woodpile templates with titanium dioxide}
\end{center}
%%%%%%%%%% Merge with supplemental materials %%%%%%%%%%
%%%%%%%%%% Prefix a "S" to all equations, figures, tables and reset the counter %%%%%%%%%%
\setcounter{equation}{0}
\setcounter{figure}{0}
\setcounter{table}{0}
\setcounter{page}{1}
\makeatletter
\renewcommand{\theequation}{S\arabic{equation}}
\renewcommand{\thefigure}{S\arabic{figure}}
\renewcommand{\bibnumfmt}[1]{[S#1]}
\renewcommand{\citenumfont}[1]{S#1}
\section{Material Characterization}
In order to confirm the removal of the polymer during the high temperature
annealing step, spectra in mid-infrared (MIR) were recorded on woodpiles
supported on CaF\textsubscript{2} before and after infiltration and
calcination using the same IR microscope also used for the NIR spectroscopy (see Methods). In Figure \ref{fig_S2}, the typical vibration bands
of the polymeric functional groups are visible (grey line). After
TiO\textsubscript{2} ALD, additionally to the characteristic signature
of the polymer, a strong broad band between 3100 and 3500 cm\textsuperscript{-1}
corresponding to the vibrational mode of OH groups ($\nu$\textsubscript{OH})
can be observed (light grey line). Dashed lines indicate the four
principal bands of the polymer, visible on both spectra, before and
after infiltration. One can notice the disappearance of nearly all
bands after calcination (black line); mainly Fabry-Pérot fringes and
a peak around 2400 cm\textsuperscript{-1}, attributed to CO\textsubscript{2},
are observed, sign of the polymer removal and TiO\textsubscript{2 }densification.
The strong CO\textsubscript{2} peak could be attributed to the polymer
decomposition; a part of the gas could indeed remain trapped in the
3D structures.
\begin{figure}[H]
\centering\includegraphics[trim=0cm 0.3cm 0cm 1cm, clip=true, width=0.75\textwidth]{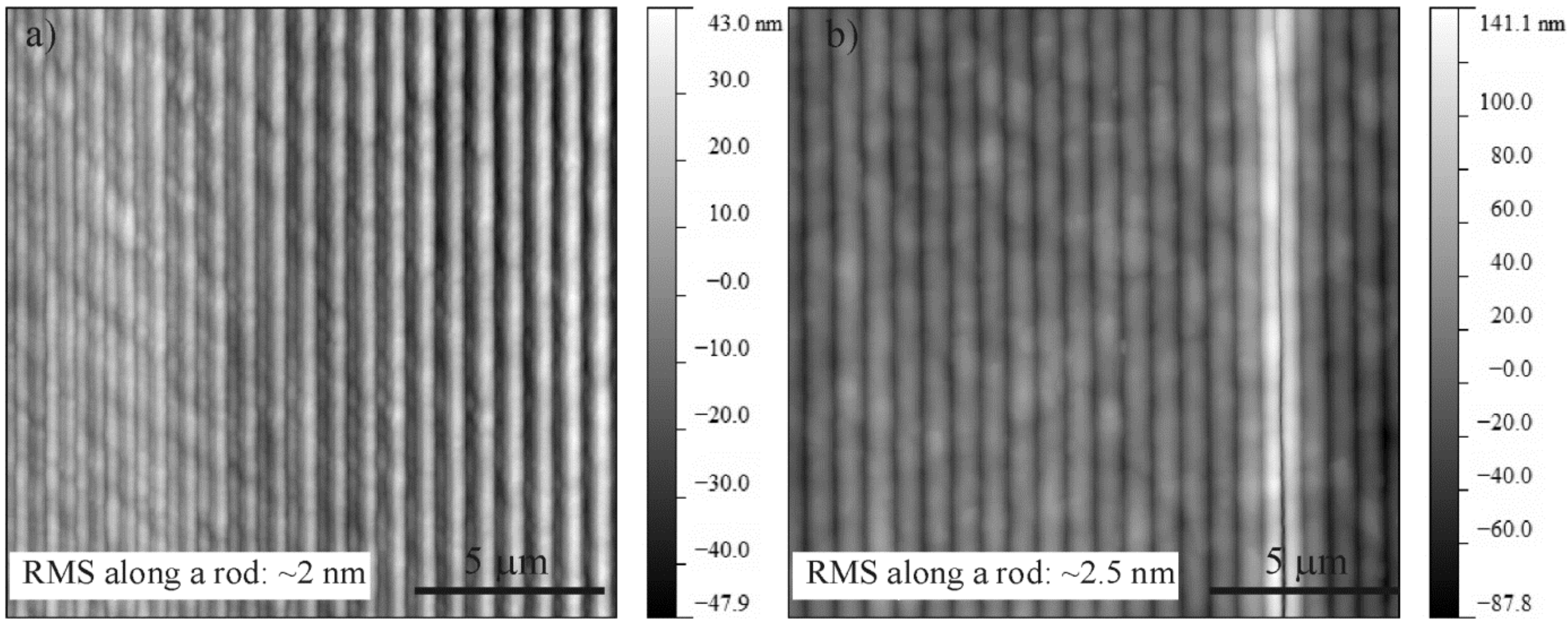} \caption{\label{fig_S1} AFM characterizations of a TiO\textsubscript{2} a)
hollow-channel and b) inverse woodpile PCs obtained after 1000 and
4500 ALD cycles and subsequent calcination.}
\end{figure}
\begin{figure}[H]
\centering\includegraphics[trim=0cm 1.5cm 0cm 0.2cm, clip=true, width=0.6\textwidth]{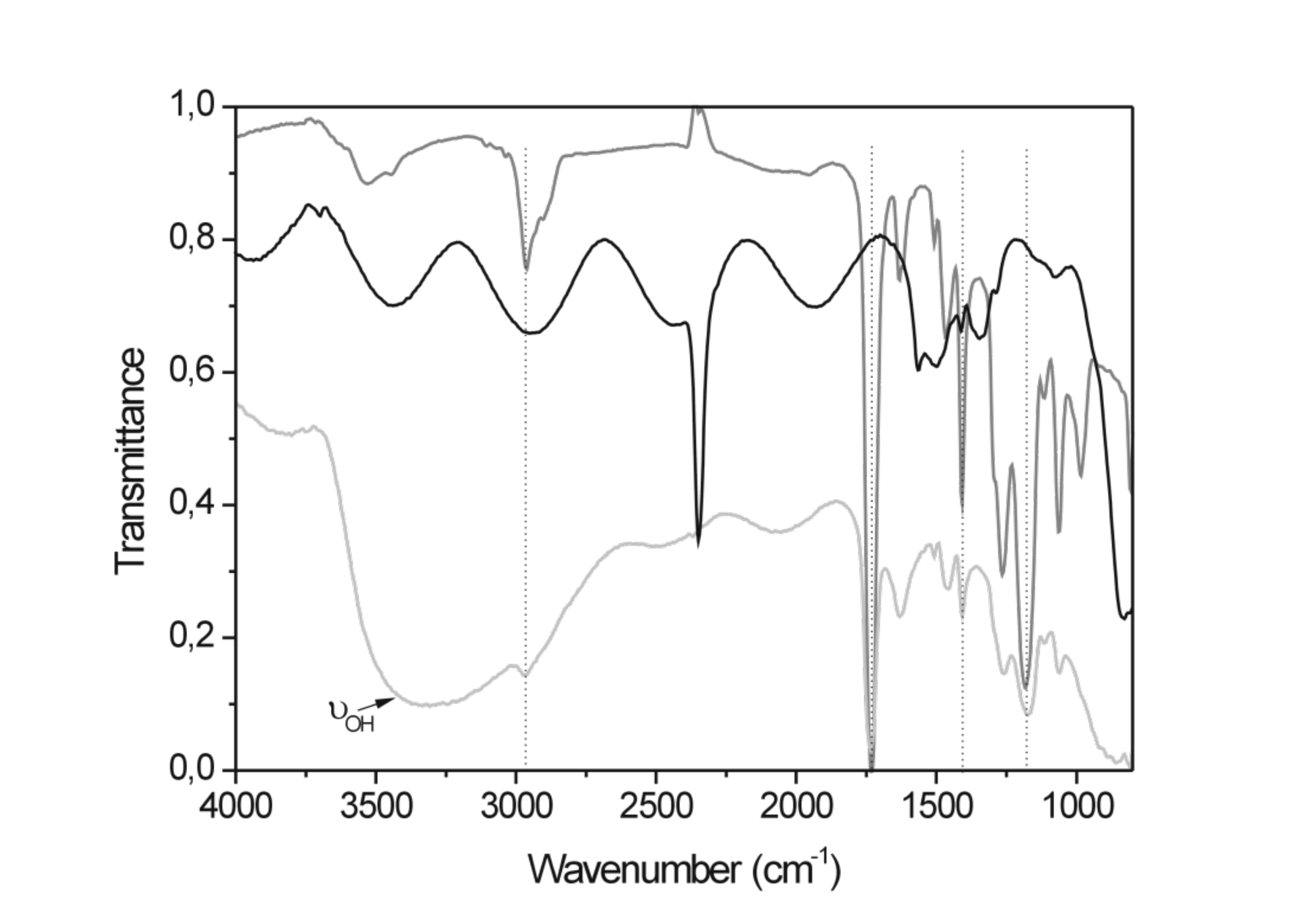} \caption{\label{fig_S2}MIR transmittance spectra are shown for polymer (grey
line), TiO\textsubscript{2}/polymer (light grey line) and TiO\textsubscript{2}
(black line) 3D structures supported on CaF\textsubscript{2} glass.}
\end{figure}
\section*{Transmittance spectra}
\begin{figure}[H]
\centering\includegraphics[trim=0cm 0.2cm 0cm 0.2cm, clip=true, width=0.6\textwidth]{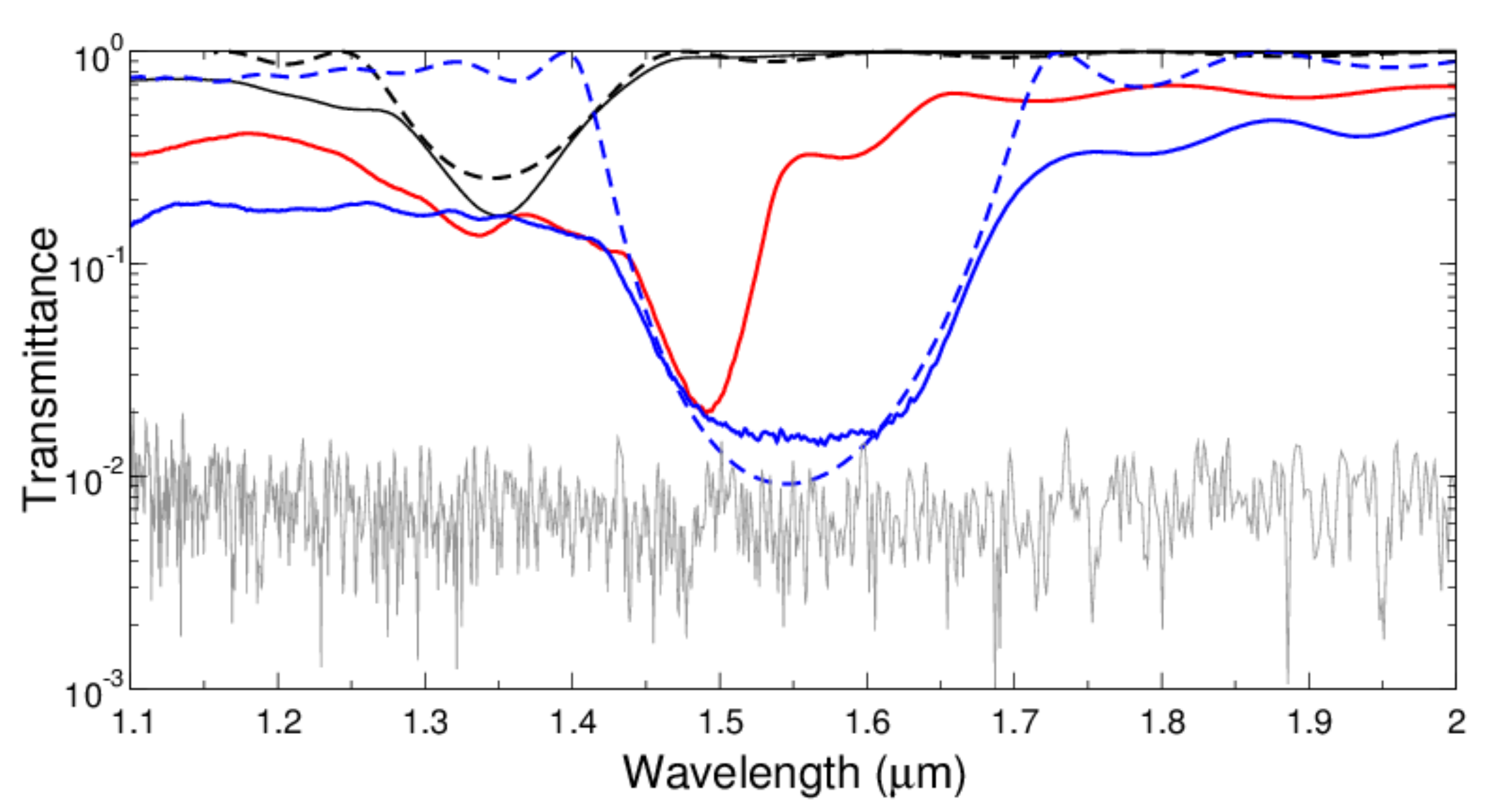} \caption{\label{fig_S3}Logarithmic representation of the data shown in Figure
2 a). Transmittance spectra of a polymer template as obtained by DLW
(black full line), the partially infiltrated TiO\textsubscript{2}
hollow-rod structures (red full line) and the TiO\textsubscript{2}
inverted structure (blue full line) showing a nearly complete bandgap.
Calculated transmittance spectra at normal incidence are shown as
dashed lines. The light gray lines denote the baseline noise indicating
the minimum measurable transmittance of the instrument, recorded on
a gold mirror.}
\end{figure}
\section*{FDTD calculations}
\begin{figure}[H]
\centering\includegraphics[width=0.6\textwidth]{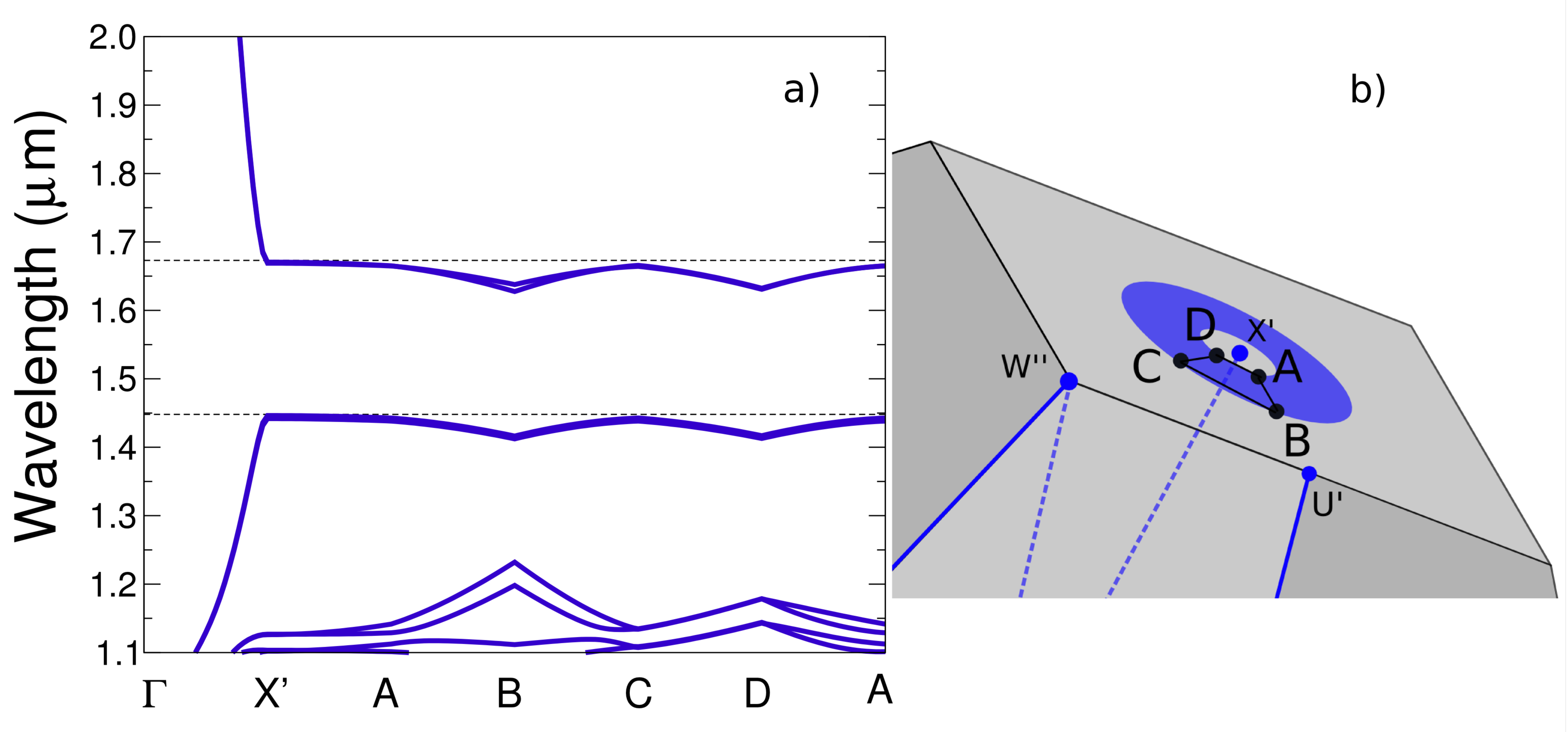} \caption{\label{fig_S4} a) Photonic band structure along a path in the first
Brillouin zone within the illumination hollow cone given by the Cassegrain
objective. The cone minimum and maximum angles are reduced according
to the refraction at the boundary between air and the host material
with to n\textsubscript{eff}=2.12 corresponding to the effective
permittivity of the TiO\textsubscript{2} with voids. In b) , a detail
of the 1\textsuperscript{st} BZ is shown together with the points
in the reciprocal space forming the path in a).}
\end{figure}

The actual measurements are not performed at normal incidence but
using a Cassegrain objective spanning angles between 10$^{\circ}$
and 30$^{\circ}$ from the normal. To verify that our calculations
at normal incidence accurately reflect the experimental situations
we calculate the band structure along a path in the boundary of the
effective illumination cone, as shown in Figure \ref{fig_S4}, after
considering the refraction in a medium of effective refractive index
n\textsubscript{eff}=2.12. As shown in the Figure S4, the $\Gamma-X'$
gap (full bandwidth=14.8 \%) and the gap found considering all the
incoming angles (full bandwidth=12.2 \%) differ only slightly in term
of width and central position. Hence, it can be concluded that simplified
normal incidence simulations provide accurate results for the transport
spectra capturing the essential characteristics of the fabricated
structures.

\end{document}